\documentclass[superscriptaddress,twocolumn,floatfx,amsmath,amssymb,prl]{revtex4-2}

\usepackage{amsmath,amssymb}
\usepackage{graphics}
\usepackage{epsfig}
\usepackage{extarrows}
\usepackage{MnSymbol}
\usepackage{mathrsfs}
\usepackage[usenames]{color}
\usepackage{appendix}
\usepackage{times}
\usepackage[colorlinks=true,linkcolor=blue,urlcolor=blue,citecolor=blue]{hyperref}

\begin{document}

\title{Universality in Ionic Three-body Systems Near an Ion-atom Feshbach Resonance}

\author{Jacek G\ifmmode\mbox{\c{e}}\else\c{e}\fi{}bala}
\affiliation{Faculty of Physics, University of Warsaw, Pasteura 5, 02-093 Warsaw, Poland}
\author{Micha{\l} Tomza}
\affiliation{Faculty of Physics, University of Warsaw, Pasteura 5, 02-093 Warsaw, Poland}
\author{Jos\'{e} P. D'Incao}
\affiliation{JILA, NIST, and the Department of Physics,
University of Colorado, Boulder, CO 80309, USA}	
\affiliation{Department of Physics, University of Massachusetts Boston, Boston, MA 02125, USA}	

\date{\today}

\begin{abstract}

We calculate the bound and scattering properties of a system of two neutral atoms and an ion near an ion-atom Feshbach resonance. Our results indicate that long-range ion-atom interactions lead to significant deviations from universal behavior derived from contact or van der Waals potentials. We find that ionic systems display an overall suppression of inelastic transitions leading to recombination rates and lifetimes of Efimov state orders of magnitude larger with respect to those for neutral atoms. We further characterize the dense spectra of triatomic molecular ions with extended lifetimes. Our results provide a deeper insight into the universality and structure of three-body ionic systems and establish them as a promising platform for exploring novel few- and many-body phenomena with long-range interactions.

\end{abstract}

\maketitle

Cold mixtures of laser-cooled ions and atoms provide exceptional opportunities for exploring and controlling quantum systems, enabling precise manipulation of both quantum states and collision dynamics~\cite{Tomza:2019,DeissNP24}. These hybrid systems have become a cornerstone of quantum science, with applications spanning quantum simulation~\cite{bissbort2013prl,jachymski2020prr}, quantum information processing~\cite{DoerkPRA10,Ratschbacher:2013}, cold controlled chemistry~\cite{Ratschbacher:2012,Sikorsky:2018,DorflerNC19,Hirzler:2022},
and few- and many-body quantum physics~\cite{CotePRL02,harter2012prl,SchurerPRL17,KleinbachPRL18}.
Recently, ultracold ion-atom mixtures have been realized in the quantum regime~\cite{feldker:2020,Weckesser:2021}, enabled by the large mass imbalance between heavy ions and light neutral atoms, which suppresses micromotion-induced heating~\cite{Cetina:2012}. This achievement allows controllable interactions near ion-atom Feshbach resonances~\cite{Weckesser:2021,Thielemann:2025,Morita:2025} through tunability of the $s$-wave scattering length~\cite{Chin:2010}, thereby opening new avenues for investigating complex quantum few- and many-body phenomena governed by long-range ion-atom interactions~\cite{casteels2011jltp,hirzler2020prr,astrakharchik2021cp,christensen2021prl,astrakharchik2021cp,Ding:2022,AstrakharchikNC23,PessoaPRL24}, as well as controlling and sympathetic cooling ions using ion-atom
confinement-induced resonances~\cite{Melezhik:2019pra,Melezhik:2021pra}.

Building on this progress, it is important to recognize that ultracold gases are intrinsically shaped by few-body processes. Among these, three-body recombination -- a fundamental chemical reaction in which three free atoms react to form a diatomic molecule carrying away the excess energy -- is the dominant mechanism driving atomic and molecular losses, thereby limiting both sample lifetimes and achievable densities~\cite{weber2003PRL} necessary for the observation of few- and many-body dynamics.
Conversely, recombination is not only a loss mechanism but also a powerful diagnostic tool. Its rate and product distribution provide sensitive probes of fundamental quantum effects such as Efimov physics~\cite{braaten:2006, Greene:2017,Naidon2017,Dincao:2018}, enable detailed studies of state-to-state ultracold chemistry~\cite{Wolf:2017, Haze:2022, Haze2023, Haze:2025, li2025pra, dorer2025ARX}, and serve as a key method for detecting Feshbach resonances in both neutral and ion-atom systems~\cite{Chin:2010,Weckesser:2021}. This multifaceted role highlights the potential that recombination has in revealing crucial aspects of few- and many-body quantum dynamics. The growing ability to control ion-atom systems at ultracold temperatures naturally motivates a closer look at how few-body processes and interactions unfold in the presence of the long-range ion-atom interactions.

Classical and semiclassical theories of ion-atom-atom systems have provided valuable insight into three-body recombination at temperatures well above the quantum regime~\cite{Perez-Rios:2014,Jesus:2015, Perez-Rios:2018,Jesus:2021, Mirahmadi:2022, Londono:2023}, establishing a solid foundation for understanding the basic dynamics. With ultracold experiments now pushing deep into the quantum domain, exciting new questions arise that naturally call for a fully quantum-mechanical, non-perturbative description. For instance, while the Efimov universality in heteronuclear, neutral systems -- characterized by short-range interactions -- is well understood \cite{Helfrich-Hammer:2010,wang2012prl}, the impact of the long-range ion-atom interaction remains largely unexplored, preventing a deeper insight into the structure and dynamics of strongly interacting ionic few-body systems~\cite{Naidon:2014l}. Equally compelling are opportunities to map out the influence of the long-range interaction on other non-universal or chemical aspects like the state distribution of diatomic molecular ions produced by recombination~\cite{Wolf:2017, Haze:2022, Haze2023, Haze:2025,li2025pra}, and the structure and stability of triatomic molecular ions, which can be potentially created in ion-atom systems.

In this Letter, we investigate universal and non-universal aspects of ionic three-body quantum systems composed of two identical bosonic atoms, $^7$Li, and a heavy ion, $^{138}$Ba$^+$. Near an ion-atom Feshbach resonance, we find that the ionic system (LiLiBa$^+$) Efimov physics is also manifested in a universal way, but with the three-body recombination rate, $L_3$, being strongly suppressed compared to its neutral counterpart (LiLiBa)~\cite{Helfrich-Hammer:2010}. We also found that the Efimov states in LiLiBa$^+$ possess lifetimes up to 5 orders of magnitude longer than those of LiLiBa, making them more accessible to experimental observation and manipulation. These results confirm that systems with long-range ion-atom interactions belong to a new class of universality~\cite{Naidon:2014l}. 
We also show that the product state distribution of recombination for both ionic and neutral systems follows the same $1/E_b$ propensity rule observed in homonuclear systems~\cite{Wolf:2017, Haze:2022, Haze2023, Haze:2025,li2025pra}, with no significant preference for forming Li$_2$ or LiBa molecules.
Finally, we characterize weakly bound triatomic molecular ions, confirming the high density of states characteristic of long-range ion-atom interactions. Together, these results provide a deeper insight on the universality, stability, and structure of ion-atom systems and establish them as a promising platform for exploring novel few- and many-body phenomena with long-range interactions.

\begin{figure}[htbp]
\includegraphics[width=\columnwidth]{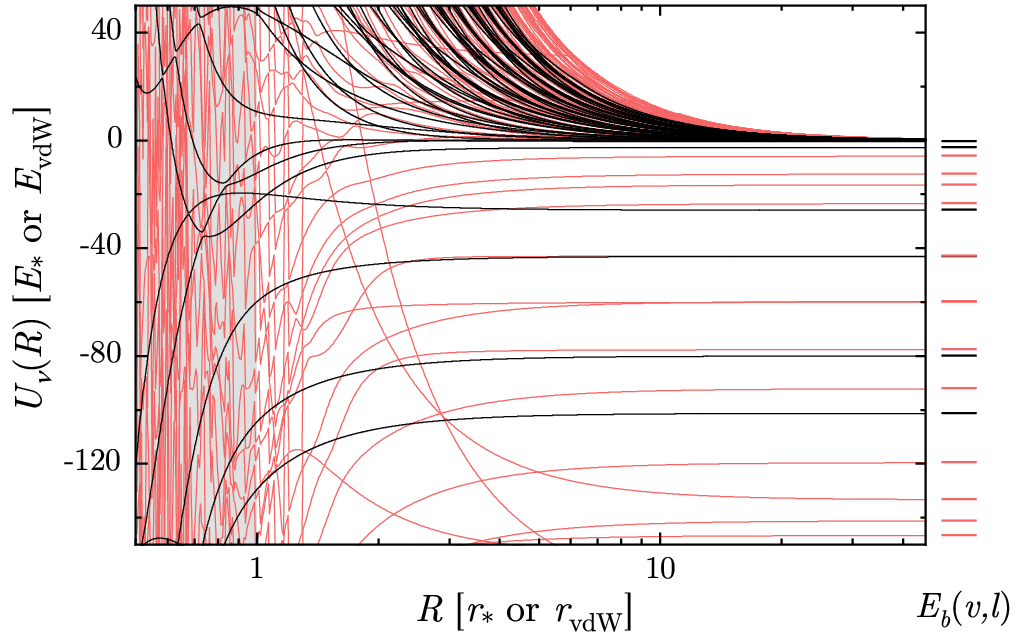}
\caption{The three-body hyperspherical potentials $U_\nu(R)$ for the LiLiBa$^+$ (black) and LiLiBa (red) systems calculated at $a_{BX} = 0.1 \ (r_* \ \text{or}  \ r_\text{vdW})$ for our interaction model supporting six $BX$ $s$-wave bound states and two $BB$ $s$-wave bound states. 
For large $R$ ($R/r_{\rm vdW} \gg 1$ or $R/r_* \gg 1$), potentials $U_\nu(R) > 0$ correspond to three-body continuum channels, describing collisions between three free atoms, while potentials $U_\nu(R) \simeq -E_b(v,l) < 0$ are atom-molecule channels describing collisions between an atom and a molecule. Here, $E_b(v,l)$ denotes the diatomic molecular binding energies of the rovibrational states of the $BX$ and $BB$ interactions. According to the number of the atom-dimer channels in the figure and considering values of $E_{\rm vdW}$ and $E_{*}$ in absolute units, we estimate that the density of diatomic states for the ionic systems to be a 100 times larger than for the neutral counterpart.}
\label{fig:1_potentials}
\end{figure}

Our investigation of the universality of the ion-atom-atom dynamics begins with the adiabatic hyperspherical representation~\cite{Suno:2002,wang2011pra,Dincao:2018}. In the hyperspherical representation, the system's internal motion and rotations are described by a set of hyperangles $\Omega$, while its overall size is described by the hyperradius $R$~\cite{Suno:2002,wang2011pra,Dincao:2018}. Bound and scattering properties of the system are obtained through solutions of the hyperradial Schr\"odinger equation
 \begin{equation}
     \left[ -\frac{\hbar^2}{2\mu} \frac{d^2}{d R^2} + U_\nu(R) - E \right] F_\nu(R) +\sum_{\nu'} W_{\nu\nu'}(R) F_{\nu'}(R) = 0 \ , \label{SchrEq}
 \end{equation}
where $\mu=(m_B^2m_X/M)^{1/2}$ is the three-body reduced mass of a system of two identical bosonic atoms, $B$, and a third dissimilar atom, $X$, of masses $m_B$ and $m_X$, respectively, total mass $M=2m_B+m_X$, and $\nu$ represents the set of quantum numbers characterizing each channel. The hyperradial Schr\"odinger equation (\ref{SchrEq}) describes the hyperradial motion governed by the three-body effective potentials $W_{\nu}(R)=U_\nu(R)+W_{\nu\nu}(R)$ supporting bound and resonant states and non-adiabatic couplings $W_{\nu\neq\nu'}(R)$ driving the inelastic transitions between the different channels. Both  $U_\nu$ and $W_{\nu\nu'}$ are determined via the solutions of the hyperangular Hamiltonian, containing all the interatomic interactions~\cite{Suno:2002,wang2011pra,Dincao:2018}. In Figure \ref{fig:1_potentials} we display the adiabatic potentials for the LiLiBa and LiLiBa$^+$ systems relevant to our present study.

Here, we assume the interactions in the three-body system to be a pairwise sum of the interatomic interactions between the identical bosons (in our case, $^7$Li), $v_{BB}$, and that between the bosons and the third particle (Ba or Ba$^+$), $v_{BX}$. In our interaction model, the interaction between identical bosons is given by the Lennard-Jones potential $v_{BB}(r)=-{C^B_{6}}/{r^6}(1-{\lambda_B^6}/{r^{6}})$ while for the interspecies interaction it is given either by a Lennard-Jones potential $v_{BX}(r)=-{C_{6}}/{r^6}(1-{\lambda_X^6}/{r^{6}})$, in the case of the neutral Li-Ba pair, or by the regularized polarization potential $v_{BX}(r)=-{C_{4}}/{r^4}(1-{\lambda_X^4}/{r^{4}})$, for the Li-Ba$^+$ pair. Here, $C_4$ and $C_6$ are long-range induction and dispersion coefficients~\cite{Tomza:2019,Derevianko:1999, 
jeziorski_moszynski_szalewicz_1994}, $r$ is the interatomic distance, and $\lambda_B$ and $\lambda_X$ parameters are adjusted to produce the desired value of the $s$-wave scattering length as well as the number of bound states each interaction pair can support.
For our studies, the interaction between Li atoms, $v_{BB}$, is set with $\lambda_B\approx 19.58\,a_0$, producing two $s$-wave Li$_2$ molecular states and the background scattering length of $-27.3\,a_0$~\cite{Moerdijk:1994,Abraham:1995,Pollack:2009L}, where $a_0$ is the Bohr radius, while we vary $\lambda_X$ for $v_{BX}$ to simulate the changes of the interspecies scattering length, $a_{BX}$, near a Feshbach resonance.

Besides the distinct short- and long-range character of the $v_{BX}$ interaction for neutral and ionic systems, another important distinction is the characteristic length and energy scales defining the onset of universality in the system.
While for neutral systems, the relevant scales are determined by the van der Waals length and energy given respectively by $r_{\rm vdW}=
(2\mu_{BX}C_{6}/\hbar^2)^{{1}/{4}}/2$ and $E_{\rm vdW}=\hbar^2/(2\mu_{BX}r_{\rm vdW}^2)$, where $\mu_{BX}=m_Bm_X/(m_B+m_X)$, for the ionic system they are defined by the polarization length and energy, $r_*=(2\mu_{BX}C_{4}/\hbar^2)^{{1}/{2}}/2$ and $E_*=\hbar^2/(2\mu_{BX}r_*^2)$, respectively. In our system, they are equal to: $r_{\rm vdW}\approx44.99 \ a_0$, $E_{\rm vdW}\approx k_B\times 6409.39\ \mu$K (or $h\times 133.55$ MHz), $r_*\approx707.03 \ a_0$, and $E_*\approx k_B\times 25.95\ \mu$K (or $h\times 0.54075$ MHz), illustrating the disparate length and energy scales relevant for neutral and ionic systems caused by the strong polarization effects. To ensure the proper comparison between neutral and ionic systems, we will express the results for each system in terms of its characteristic length and energy scales whenever appropriate.

\begin{figure*}[htbp]
\includegraphics[width=\textwidth]{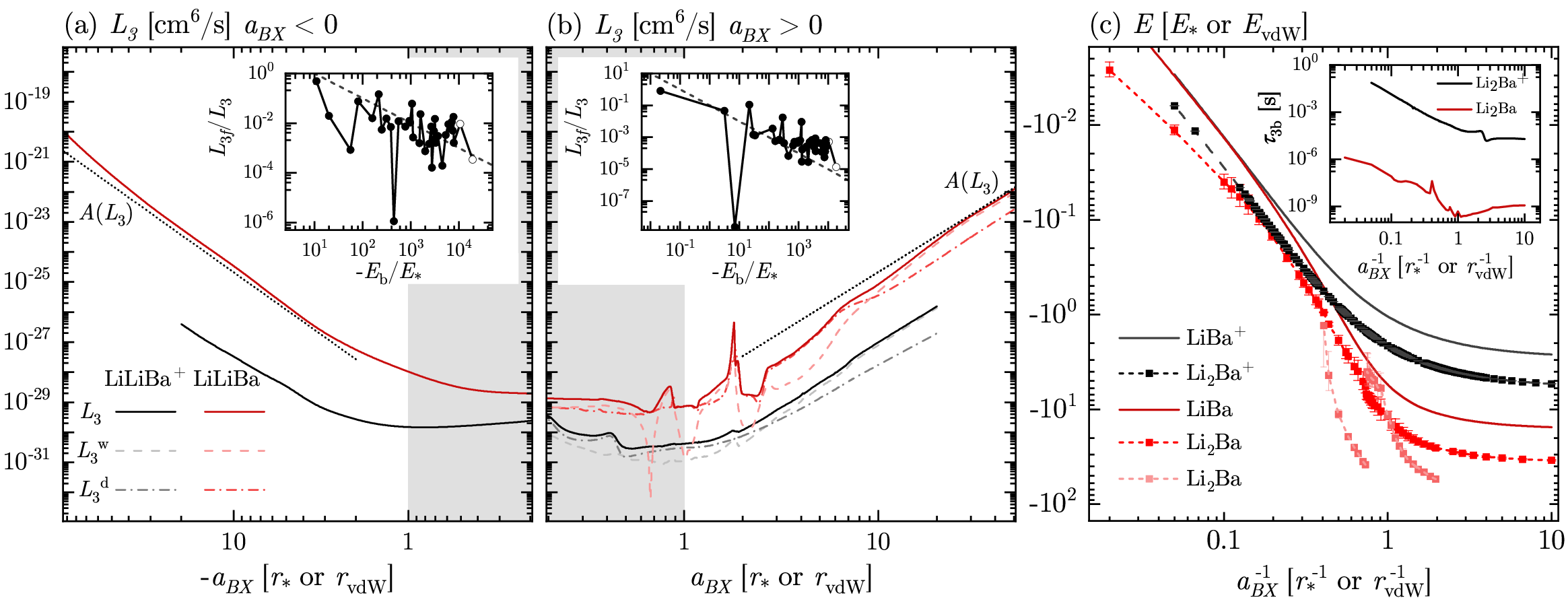}
\caption{(a), (b): Three-body recombination rate $L_3$ and partial rates $L_3^{\rm w}$ and $L_3^{\rm d}$ for LiLiBa (red) and LiLiBa$^+$ (black). The results for LiLiBa$^+$ recombination are multiplied by $(r_\text{vdW}/r_*)^4$ in order to properly compare that with recombination of LiLiBa systems. For small values of $|a_{BX}|\lesssim r_*$ (or $r_{\rm vdW}$), $L_3$ display resonant effects associated with high-partial waves diatomic molecular states~\cite{wang2012pra}, some of which are not resolved in the figure. Dotted lines represent the amplitude for $L_3$, $A(L_3)$, from the universal theory~\cite{Helfrich-Hammer:2010}. The insets of panels (a) and (b) display the product state distribution of LiLiBa$^+$ recombination, $L_{3f}/L_3$ [see Eq.~(\ref{L3})], in terms of the molecular final state binding energy, $E_b$, displaying the $1/E_b$ propensity rule of Ref.~\cite{Haze2023}. (c) Energy of the lowest Efimov state for LiLiBa (red) and LiLiBa$^+$ (black) and the corresponding width $\Gamma$ expressed as error bars. Two additional trimer states associated with two-body rotational states (pink) \cite{wang2012pra} are also presented, together with their energies and widths. The inset in (c) shows the lifetime $\tau=\hbar/\Gamma$ of the Efimov states for both LiLiBa and LiLiBa$^+$.}
\label{fig:2_merged}
\end{figure*}

For our scattering calculations, we define the three-body recombination constant for a system of two identical bosons at collision energy $E$ as~\cite{mehta2009prl},
\begin{equation}
    L_3(E) = \sum_{J}\sum_{f,i}32 \pi^2 \hbar \frac{(2J+1)}{\mu k^4} |S^{J}_{fi}|^2=\sum_f L_{3f} \ ,\label{L3}
\end{equation}
with $k=(2\mu E/\hbar^2)^{1/2}$, $f$ running over all final (atom-dimer) channels, $i$ over the initial (three-body continuum) channels (see Fig.~\ref{fig:1_potentials}). Here, $L_{3f}$ is the partial recombination rate into the final state $f$, and the corresponding $S$-matrix obtained from the solutions of Eq.~(\ref{SchrEq}) using the methodology developed in Ref.~\cite{wang2011pra}. For the regime of ultracold collisions ($E\ll E_{\rm vdW}$ or $E_*$), we only consider the lowest total three-body angular momentum $J=0$ as higher partial-waves contributions are suppressed in this regime~\cite{esry2001pra,dincao2005prl}.

Figures~\ref{fig:2_merged}(a) and~\ref{fig:2_merged}(b) present the three-body recombination rate, $L_3$,
for LiLiBa (red solid lines) and LiLiBa$^+$ (black solid lines) at collision energy $E/k_B=0.01\mu$K as a function of the interspecies scattering lengths, $a_{BX}$. 
In the figure, we also show the rate $L_3^{\rm w}$ for recombination into the weakly bound Feshbach molecular state ($a_{BX}>0$) as well as recombination into all other deeply bound states, 
$L_3^{\rm d}=L_3-L_3^{\rm w}$. For the ionic system, we use a Li-Li interaction supporting 2 $s$-wave states while the Li-Ba$^+$ supports 6 $s$-wave states (for a total of $\sim$10 molecular states). For the neutral system, we use for both Li-Li and Li-Ba interactions that support 2 $s$-wave states (for a total of $\sim$40 states). For both ionic and neutral systems, we include in our calculations 50 three-body continuum channels. 
We estimate our results to be converged within 1-2\% level.
{In the attempt of gaining a better understanding of how recombination proceeds for ionic systems, in the inset of Figs.~\ref{fig:2_merged}(a) and~\ref{fig:2_merged}(b) we show the product state distribution $L_{3f}/L_3$ in terms of the binding energy of the final molecular state, $E_b$, for both $a_{BX}>0$ and $a_{BX}<0$, respectively.}
Although for LiLiBa$^+$ there are two types of molecular final products (Li$_2$+Ba$^+$ or LiBa$^+$+Li), our results in the inset of Figs.~\ref{fig:2_merged}(a) and~\ref{fig:2_merged}(b) verifies the same $L_{3f}/L_3\sim1/E_b$ propensity rule found for neutral homonuclear $^{87}$Rb and $^{85}$Rb recombination~\cite{Wolf:2017, Haze:2022, Haze2023, Haze:2025,li2025pra}, with no preference over Li$_2$ (open symbol) or LiBa$^+$ (closed symbol) molecular states. This indicates that the same physical processes found for neutral (homonuclear) systems apply to ionic (heteronuclear) systems despite the highly complex nature of the three-body interactions at short distances ($R\lesssim r_*$  or $R\lesssim r_{\rm vdW}$ in Fig.~\ref{fig:1_potentials}).

For large values of $|a_{BX}|$, $|a_{BX}|\gtrsim r_*$ or $r_{\rm vdW}$, Fig.~\ref{fig:2_merged} shows that $L_3$ follows the expected $a_{BX}^4$ scaling behavior~\cite{Dincao:2018}. Moreover, in this range of $a_{BX}$, interference and resonant behaviors are expected. They are associated to the $n$-th ($n=0,1,2,\ldots$) Efimov state for $a_{BX}=a_+ e^{n\pi/s_0}>0$ and $a_{BX}=a_- e^{n\pi/s_0}<0$, respectively~\cite{braaten:2006,Greene:2017,Naidon2017,Dincao:2018}, with $s_0$ being the Efimov universal coefficient controlling the strength of the Efimov interaction. As usual, the expected values of three-body parameters $|a_+|$ and $|a_-|$ are typically larger than the characteristic range of the interactions~\cite{Wang:2012}. However, since LiLiBa$^+$ and LiLiBa are both unfavorable mass-imbalanced systems ($m_X/m_B\gg1$), the Efimov coefficient is small, $s_0\approx0.03562$, and geometric scaling extremely large, $e^{\pi/s_0}\approx2\times10^{38}$, in comparison to $s_0\approx1.00624$ and $e^{\pi/s_0}\approx22.7$ for three identical bosons systems. As a result, it is unlikely that such systems will display interference and resonance phenomena for experimentally accessible values of $a_{BX}$. Nevertheless, universal behavior is still expected for the amplitude of $L_3$, $L_3^{\rm w}$, and $L_3^{\rm d}$ as shown in Ref.~\cite{Helfrich-Hammer:2010}. Such universal results for $L_3$ (see End Matter) depend on the mass ratio $m_X/m_B$, $a_+$, $a_-$, and $\eta$, the non-universal inelasticity parameter describing decay to deeply bound molecular states~\cite{braaten:2006,Dincao:2018}. For the LiLiBa neutral system, our numerical calculations for $L_3$ agree with the universal results of Ref.~\cite{Helfrich-Hammer:2010}. The straight lines in Figs.~\ref{fig:2_merged}(a) and~\ref{fig:2_merged}(b) represent the amplitude of the universal $L_3$ without the interference and resonant terms (see End Matter) and are denoted by $A(L_3)$. From the comparison to the universal results we obtain $a_-\approx-200 \,r_{\rm vdW}$, $\eta\approx 0.17$ for $a_{BX}<0$ and $\eta\approx0.027$ for $a_{BX}>0$. 

For LiLiBa$^+$, we could not identify any reasonable set of parameters $a_+$, $a_-$, and $\eta$ within the framework of the universal theory~\cite{Helfrich-Hammer:2010} that reproduces the recombination amplitude shown in Fig.~\ref{fig:2_merged}. Furthermore, by employing different interaction models for the Li-Ba$^+$ potential -- each supporting a different number of bound states -- we confirm that our results for the LiLiBa$^+$ system are themselves universal, yet suppressed by a factor of about 250 compared to the corresponding universal predictions of Ref.~\cite{Helfrich-Hammer:2010}. We attribute this suppression of inelastic transitions to the difference on how the LiLiBa$^+$ interactions change as $R$ approaches short-distances, as compared to the LiLiBa case. As shown in Fig.~\ref{fig:1_potentials}, the three-body potentials for LiLiBa$^+$ vary more slowly as approaching short-distances than those for LiLiBa, leading to smaller non-adiabatic couplings $W_{\nu\nu'}$ in Eq.~(\ref{SchrEq}) and suppressing inelastic transitions. We have verified this reduction of non-adiabatic couplings numerically.

Evidently, these findings raise the question of what makes the universality for ionic systems different than that of neutral atoms. In fact, as shown in Ref.~\cite{Naidon:2014l}, systems interacting via $1/r^6$ and $1/r^4$ interactions (in our case, the LiLiBa and LiLiBa$^+$ systems, respectively) belong to different universality classes and are characterized by different values of the universal three-body parameters (e.g., $a_-$)~\cite{Comment01}. 
Although the universal theory~\cite{Helfrich-Hammer:2010} -- constructed within the framework of the effective-field theory (EFT) assuming contact, zero-range interactions -- does not provide the values of the three-body parameters, they provide universal expressions for $L_3$ in terms of such three-body parameters, as well as various universal relationships between them. The validity of these expressions has been extensively verified for neutral atomic systems belonging to the $1/r^6$ universality class. Nevertheless, systems within the $1/r^4$ universality class have fundamentally different low-energy properties manifested, for instance, via the effective-range expansion~\cite{flambaum1998pra,HRSadeghpour_2000,OMalley:1962,SJBuckman_1989}. We provide a detailed analysis of the difference between the effective range expansion for both $1/r^6$ and $1/r^4$ interactions in the End Matter. Such modifications of the low-energy behavior suggest that changes to the universal EFT are required in order to properly describe the universality of ionic three-body systems, similar to those in Refs.~\cite{Odell:2021,Odell:2023}, which demonstrate the importance of treating $1/r^6$ and $1/r^4$ interactions within the EFT framework. Our results on the changes observed in our $L_3$ calculations for neutral and ionic systems, along with the $1/r^4$ universality class studies of Ref.~\cite{Naidon:2014l}, indicate that the ionic nature of the two-body interaction affects not only the universality of the three-body parameter but also the universal collision rates and other universal relationships derived from the EFT framework~\cite{Helfrich-Hammer:2010}. We provide a detailed analysis of the mechanisms of the ion-atom-atom universality in the Supplemental Material~\cite{supplemental}.
\nocite{dincao2005pra,wang2011praX,Hafner:2017PRA,Oi:2024PRA}

\begin{figure}[t!]
\includegraphics[width=\columnwidth]{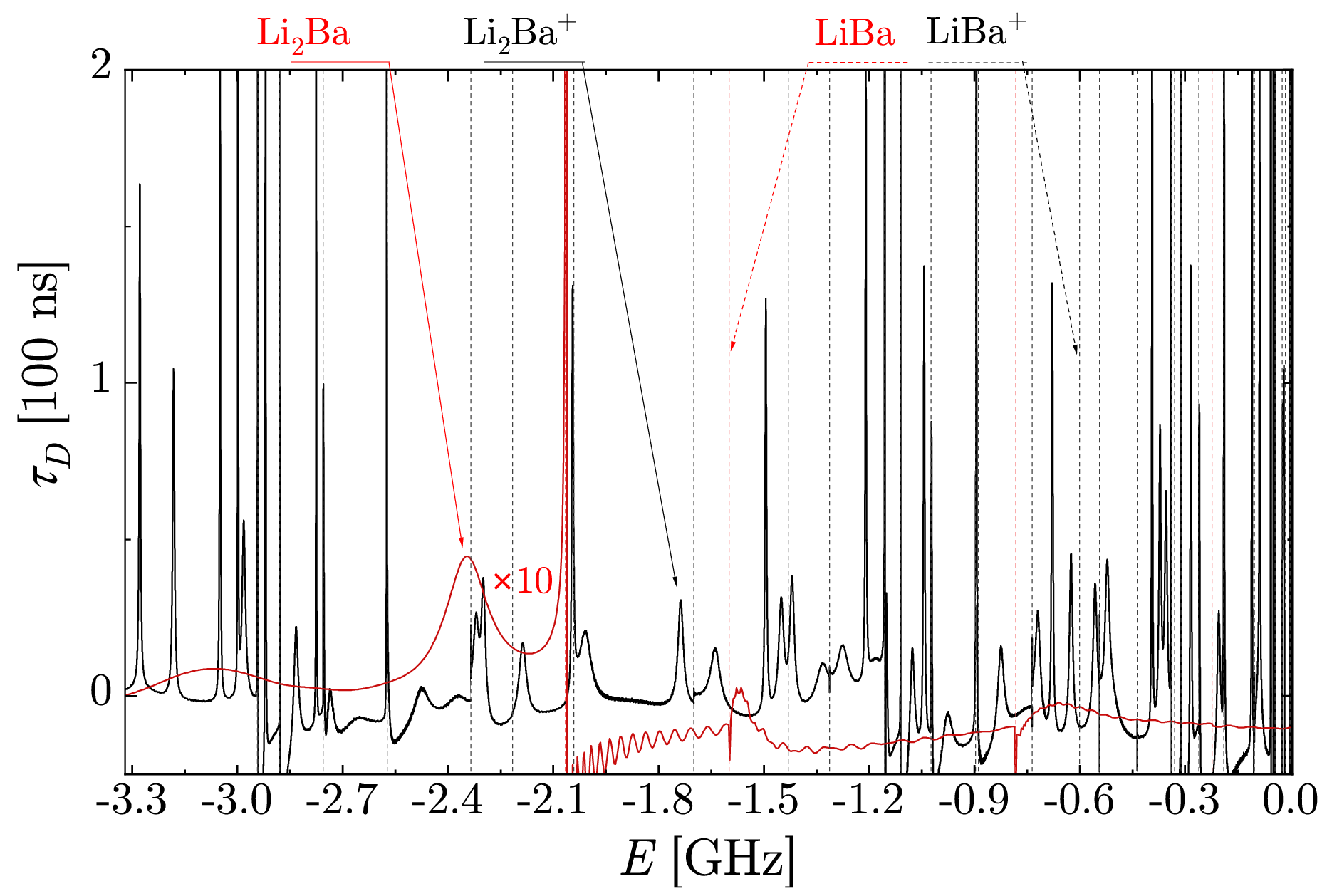}
\caption{The time delay for $^7\text{Li}^7\text{Li}^{138}\text{Ba}$ (red) and $^7\text{Li}^7\text{Li}^{138}\text{Ba}^+$ (black) calculated for $a_{BX} = 0.1 \ (r_* \ \text{or} \ r_\text{vdW})$. The results for the neutral system have been multiplied by a factor of 10. The vertical dashed lines are the energies of two-body molecular states to which the effective potentials $W_\nu(R)$ converge at large values of $R$. The peaks and width of the time-delay parameter describe the energies and lifetimes of the three-body bound states supported by $W_\nu(R)$.
}
\label{fig:3_time_delay}
\end{figure}

Although both the LiLiBa and LiLiBa$^+$ systems are unfavorable for Efimov physics (due to the value of the Efimov coefficient, $s_0\approx0.03562$) our calculations in Fig.~\ref{fig:2_merged}(c) show that for $a_{BX}>0$ the Efimov ground state remains bound for values of $a_{BX}/r_{\rm vdW}$ and $a_{BX}/r_*$ much smaller than $e^{\pi/s_0}\approx2\times10^{38}$. This can be explained by the variational principle of Ref.~\cite{bruch1973pr}, which states that the ground state binding energy of a triatomic molecule can not exceed three times that of the diatomic molecule, preventing the Efimov ground state from crossing the atom-dimer threshold and becoming unbound~\cite{lee2007pra,mestrom2017pra}. In Figure~\ref{fig:2_merged}(c), we show the diatomic molecular energies of the corresponding Feshbach states, $E^*_{\rm 2b}$ (solid lines), along with the energies $E_{\rm 3b}$ (dashed lines with symbols) and width $\Gamma_{\rm 3b}$ (error bars) of the triatomic molecular states obtained via the time-delay calculations similar to that of Ref.~\cite{Nielsen:2002}. While the energies of the Efimov ground state for the LiLiBa and LiLiBa$^+$ systems are similar (in relative units), their corresponding lifetime $\tau_{\rm 3b}=\hbar/\Gamma_{\rm 3b}$ are substantially different. The Efimov LiLiBa$^+$ ground state has a lifetime around 5 orders of magnitude longer than those of LiLiBa [see the inset of Fig.~\ref{fig:2_merged}(c)], with lifetimes as long as 100 ms for the range of $a_{BX}$ shown in Fig.~\ref{fig:2_merged}(c). 
The longer lifetimes for LiLiBa$^+$ states can be understood by the combination of their more weakly bound nature compared to the LiLiBa states -- determined by the ratio of characteristic energies scales, $E_*/E_{\rm vdW}\approx4.05\times10^{-3}$) -- and the suppression by a factor of 250 of the inelastic transitions for LiLiBa$^+$ [see Figs.~\ref{fig:2_merged}(a) and~\ref{fig:2_merged}(b)] thus leading to an overall factor of $E_*/E_{\rm vdW}/250\approx1.62\times10^{-5}$.
Long-lived ground Efimov states in neutral atoms have so far remained elusive, making ionic Efimov states likely to produce interesting regimes in the ultracold ion-atom gas mixture. While ionic Efimov states are extremely weakly bound ($|E_{\rm 3b}|\ll E_*\approx25.95 \ \mu$K for LiLiBa$^+$), and difficult to populate due to the yet limited temperature experiments can reach, their extreme large extent, $\langle R\rangle_{\rm 3b}\gg r_*$, can exceed the average interatomic distances (approximately 30 $r_*$ for densities around $10^{12}$ cm$^{-3}$) potentially leading to novel physical regimes~\cite{CotePRL02,casteels2011jltp, astrakharchik2021cp,christensen2021prl,jachymski2020prr,hirzler2020prr,doerk2010prl,gerritsma2012prl,bissbort2013prl}.

To further explore binding phenomena in ionic few-body states, and the differences to neutral atom systems, we also calculated the energy spectrum and the corresponding lifetimes of the triatomic molecular states beyond the energy range relevant for Efimov physics, i.e., $E>E_{\rm vdW}\approx h\times133.55$ MHz for LiLiBa and $E>E_*\approx h\times0.54075$ MHz for LiLiBa$^+$. Figure~\ref{fig:3_time_delay} shows our calculations of the time delay $\tau_D(E)$~\cite{Nielsen:2002} for a broad range of energies (from 0 to $h\times3.3$ GHz), with molecular energies $E_r$ determined through the values of $E$ in which $\tau_D(E)$ is maximal, i.e., $d\tau_D(E)/dE=0$, and corresponding lifetimes given by $\tau_r=\tau_D(E_r)/4$. Figure~\ref{fig:3_time_delay} shows that, similar to the diatomic case (see caption of Fig.~\ref{fig:1_potentials}), the energy spectrum of triatomic molecular ions (black curve) is also much denser than that of the neutral system (red curve). [Note that in Fig.~\ref{fig:3_time_delay} the vertical dashed lines indicate the energy of the LiBa (red) and LiBa$^+$ (black) molecular states.] 

Our calculations in Fig.~\ref{fig:3_time_delay} show that the Li$_2$Ba three-body molecular states have significantly shorter lifetimes than typical Li$_2$Ba$^+$ molecular states. Notably, certain states in the ionic system are unusually narrow. Such states are characterized by slightly increased lifetimes for lower binding energies, ranging from $1$ to $10 \ \mu$s.

In summary, we investigated universal and non-universal aspects of ionic three-body systems composed of two identical bosonic atoms, $^7$Li, and a heavy ion, $^{138}$Ba$^+$. We found that near an ion-atom Feshbach resonance, the ionic system (LiLiBa$^+$) exhibits Efimov physics in close analogy to its neutral counterpart (LiLiBa), but with recombination rate, $L_3$, strongly suppressed compared to the neutral case~\cite{Helfrich-Hammer:2010}, suggesting that the long-range ion-atom interaction defines a class of universality where both universal three-body parameters~\cite{Naidon:2014l} and their universal relationships are distinct from universality for neutral atom systems~\cite{Helfrich-Hammer:2010}.
We characterized the energy and lifetime of Efimov ground state in both LiLiBa$^+$ and LiLiBa systems and found that the lifetime of the ionic Efimov state is five orders of magnitude longer than its neutral counterpart, consistent with the difference of characteristic energy scales ($E_{\rm vdW}$ and $E_*$) and the suppression factor found for the ionic $L_3$. 
We also characterized weakly bound triatomic molecular ions, confirming the expected high density of states characteristic {of} long-range ion-atom interactions. The generally longer lifetimes of ion-atom-atom systems constitute a key advantage for future experiments, opening new avenues to explore few- and many-body effects in ultracold ion-atom systems.

An interesting pathway for future research involves the description of the  dependence of universality for ionic systems on the narrowness of 
Feshbach resonance. 
The departure from the usual universal behavior in Efimov physics near narrow Feshbach resonances stems from the appearance of a new, large, two-body length scale associated to the effective range, which leads to modifications of the long-range behavior of the Efimov interactions 
\cite{petrov2004prl,gogolin2008prl,Schmidt:2012,wang2011praY,yudkin-2024,Langmack:2018PRA}. Taking into account the relevant length scales, $r_*$, we do expect similar effects in ionic systems. 

\begin{acknowledgments}
The authors gratefully acknowledge the National Science Centre Poland (grants No.~2020/38/E/ST2/00564 and 2023/49/N/ST2/03439) for financial support, the Poland's high-performance computing infrastructure PLGrid (HPC Centers: ACK Cyfronet AGH) for providing computer facilities and support (computational grant No.~PLG/2024/017527), and Hans-Werner Hammer for fruitful discussions. J. P. D. also acknowledges partial support from the U.S. National Science Foundation, Grant No.~PHY-2452751, and NASA/JPL 1502690.
This research was supported in part by the National Science Foundation under Grant No.~NSF PHY-1748958 to the Kavli Institute for Theoretical Physics (KITP).
\end{acknowledgments}

\begin{center}
    \textit{Data availability—The data that support the findings of
 this article are openly available~\cite{DAS:1_data}.} 
\end{center}

\section{End matter}

\subsection{Effective range theory}

Effective range theory is an important tool for the analysis of low-energy scattering. It became an essential framework for development of the universal theory of ultracold collisions. Effective range theory provides the leading terms in the expansion of $k^{2l+1} \cot{\delta_l}$ as a function of the collision energy, where $\delta_l$ is the phase shift of the scattering wavefunction for the angular momentum $l$, and $k$ is the wavenumber. For collision energies approaching the $s$-wave regime, we consider only $l=0$. In this regime, the shape-independent approximation to effective range expansion becomes
\begin{equation}
k \cot{\delta_0} = -\frac{1}{a} + \frac12 r_\text{eff}k^2 + \mathcal{O}(k^4) \ ,
\label{eq:neutralERE}
\end{equation}
which is valid for $k < {r_\text{eff}}^{-1}$, where $r_\text{eff}$ is the effective range, and where $a$ is the two-body scattering length. Eq.~\eqref{eq:neutralERE} is valid only for short-range potentials, which fall of faster than any power of $1/r$. The terms of the expansion are modified for inverse power-law potentials, i.e., $v(r) \rightarrow -C_n/r^n , \ r \rightarrow \infty$. Nevertheless, if the potentials decay sufficiently rapidly at large $r$, only the higher order terms are modified. For example, the term $\mathcal{O}(k^4)$ changes to $\mathcal{O}(k^4 \ln(k))$ for $n = 6$~\cite{Calle:2010_ERT_VDW, Levy:1963_ERT_VDW}.

The form of solutions of the free and regular radial Schr\"odinger equation at $k=0$ allows us to approximate the effective range with an energy-independent expression
\begin{equation}
r_\text{eff}(a) = 2\int_0^\infty [v_0^2(r) - u_0^2(r)] \ \text{d}r \ ,
\label{eq:ert_approx}
\end{equation}
where $v_0(r)=1-r/a$ is the solution to the free radial Schr\"odinger equation, and $u_0(r)$ -- the solution to Schr\"odinger equation with a finite potential $V(r)$~\cite{flambaum1998pra,OMalley:1962}.
However, Eq.~\eqref{eq:ert_approx} can only be obtained by approximating the solutions at finite $k$ to $v_0$ and $u_0$ at short-range and setting the value of the integral to 0 at long-range, where the two-body potential effectively vanishes. This assumption is valid only for a certain class of long-range two-body potentials of the form $V(r) \rightarrow -C_n/r^n  , \ r \rightarrow \infty$, where $n > 4$.

Long-range potentials, like the ones representing the ion-atom induction dominating interaction at $n=4$, require a more general approach, which introduces terms proportional to $\sim$${k}$ and $\sim$${k}^2 \ln{k}$ in the $k \cot{\delta_0}$ expansion~\cite{OMalley:1960}
\begin{equation}
k \cot{\delta_0} = -\frac{1}{a} + c_1^*(a) k   + \frac{16 r_*^2}{3a}  k^2\ln{\left(\frac{r_* k}{2}\right)} + \frac12 R_{\rm eff}^*(a) k^2 + \mathcal{O}(k^3) \ ,
\label{eq:ionERE}
\end{equation}
where $c_1^*(a) =  {4\pi r_*^2}/(3a^2)$, and where
\begin{equation}
 R_{\rm eff}^*(a) = r_\text{eff}^* + \frac{4 \pi r_*}{3} + \frac{160 r_*^2}{9a} - \frac{64 \psi\left(\frac32\right) r_*^2}{3a} - \frac{16\pi r_*^3}{3a^2} - \frac{32 \pi^2 r_*^4}{9a^3} \ , 
\label{eq:coeffk2}
\end{equation}
is the \textit{generalized} effective range for the ion-atom interaction. In the above equation, we define $r_{\rm eff}^*$ to be the ion-atom effective range.
We relate the modifications in the ionic $k \cot \delta_0$ expansion to corresponding changes in the universal theory of few-body collisions arising from long-range interactions.

\begin{figure}[t!]
\includegraphics[width=\columnwidth]{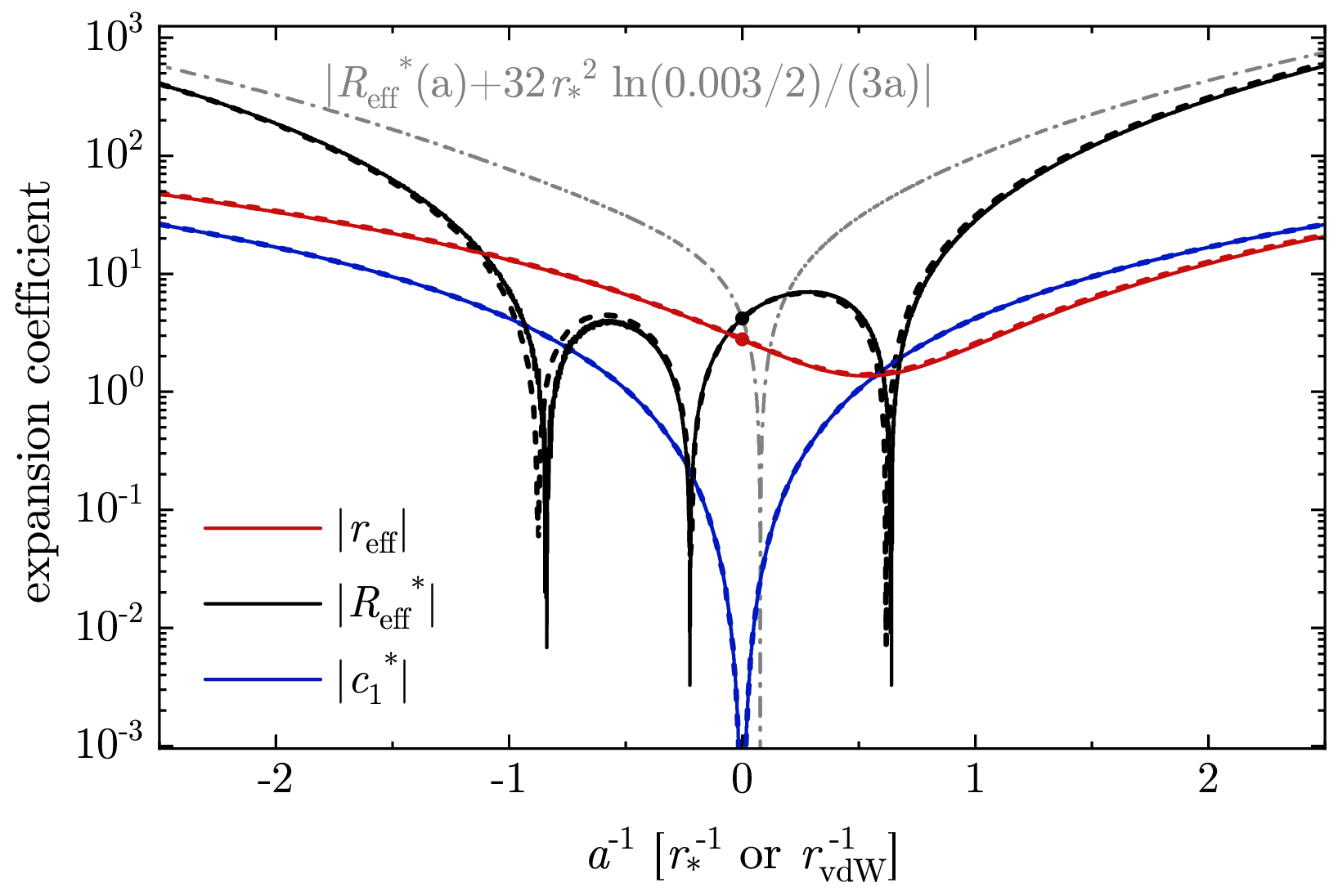}
\caption{The (generalized) effective ranges for the ion-atom (black solid line) and neutral (red solid line) two-body interactions. The coefficient $\sim$${k}$ obtained for the ionic expansion is marked as $c_1^*$ (blue solid line). Dashed lines represent the analytical expressions given by Eq.~(34) of Ref.~\cite{Chin:2010} for the neutral interaction (red), the expression for $c_1^*(a)$ (blue), and the expression given by Eq.~\eqref{eq:coeffk2}, in which we set $r_{\rm eff}^* = 0$ (black) -- for the ion-atom interaction. The gray dot-dashed line shows the generalized ionic effective range plus the estimated logarithmic term at $k = 0.003 \ r_{*}^{-1}$.}
\label{fig:sm_ert}
\end{figure}

We present the calculated results for the $k \cot{\delta_0}$ expansion in Fig.~\ref{fig:sm_ert}. For neutral and ion-atom interactions, we fit the expression in Eq.~\eqref{eq:neutralERE} and Eq.~\eqref{eq:ionERE}, respectively, to numerically obtained values of $k \cot \delta_0$. The coefficients of the effective range expansion are plotted against $1/a$ in units rescaled for both the neutral and ion-atom interactions. 
The results for the neutral $\sim$${k}^2$ coefficient, i.e., $r_{\rm{eff}}$, agree with the analytical expression reported in Ref.~\cite{Chin:2010}. Likewise, the ionic $\sim$${k}$ coefficient accurately reproduces the expected $1/a^2$ dependence. For the ion-atom system, the $\sim$$k^2$ coefficient is extracted by subtracting the logarithmic term from $k \cot{\delta_0}$ prior to fitting. Specifically, we evaluate the $\sim$$\ln k$ contribution at $k = 0.003\, r_*^{-1}$. At this value, $k \cot{\delta_0}$ converges to $-1/a$, confirming the correct threshold behavior. The numerical values are compared to Eq.~\eqref{eq:coeffk2} under the approximation $r_{\rm eff}^* \approx 0$. Any deviations from this formula provide an estimate of the ion-atom effective range. Because the non-linear $\sim$$\ln k^2$ term reduces the numerical stability of the fit, we assess its influence by adding it to the $\sim$$k^2$ coefficient after the fitting procedure. This term has a significant impact on the $\sim$$k^2$ behavior of the $k \cot{\delta_0}$ function.

\subsection{Three-body recombination for $a_{BX} > 0$}
The LiLiBa system has a large mass imbalance measured by $\delta = {m_X}/{m_B}$, where $m_X$ is the mass of one dissimilar atom $X$, in our case the Barium atom, while the two Lithium atoms have masses $m_B$. We modify the implementation of the universal theory to the three-body problem for large scattering lengths to account for the heteronuclear properties of our system. Following Ref.~\cite{Helfrich-Hammer:2010}, the rate $L_3^{\rm w}$ for recombination into the weakly bound 
Feshbach molecular state ($a_{BX}>0$) is given by 
\begin{equation}
     L_3^{\rm w} =2 C(\delta) \frac{D \left( \sin^2[s_0 \ln(a_{BX}/a_{+})] + \sinh^2(\eta) \right)}{\sinh^2(\pi s_0 + \eta) + \cos^2[s_0 \ln(a_{BX}/a_{+})]} \frac{\hbar a_{BX}^4}{m_X} \ ,
      \label{eq:alpha_w}
\end{equation}
where $D = 128\pi^2(4\pi-3\sqrt{3})$, $a_{+}$ is the three-body parameter ($a_{+} > 0$), $\eta$ is the inelasticity parameter, where $\eta \ll 1$ means low decay probability into deep dimers and $\eta \gg 1$ high decay probability, and the mass-dependent coefficient is denoted by $C(\delta)$.
We present results for three-body recombination rates $L_3$ rather than the event rate constant $\alpha_{\rm w} = L_3/2$, which takes into account that two identical atoms are taking part in the collision. 
Ref.~\cite{Helfrich-Hammer:2010}  provides the analytical formula for the mass coefficient
\begin{equation}
C(\delta) = \frac{(1+\delta)^2 \arcsin[1/(1+\delta) ]-\sqrt{\delta(2+\delta)}}{2(4\pi-3\sqrt{3})} \ ,
\end{equation}
which is valid for $\delta >2$. The universal rate for recombination into deep dimers is given by
 \begin{equation}
      L_3^{\rm d} = 2 C(\delta) \frac{D  \coth(\pi s_0) \cosh(\eta) \sinh(\eta) }{\sinh^2(\pi s_0 + \eta) + \cos^2[s_0 \ln(a_{BX}/a_{+})]} \frac{\hbar a_{BX}^4}{m_X} \ .
      \label{eq:alpha_d}
 \end{equation}
For $a_{BX} > 0$, the Efimov state is expected to unbind into the Feshbach dimer at $a_{+}$, while successive Efimov levels cross the atom-dimer threshold at scattering lengths that are separated by the universal factor $e^{\pi / s_0}$. The interference between the decay pathways produces the characteristic log-periodic suppression of the three-body recombination rate, which is described by trigonometric factors $\sim$$\sin^2[s_0 \ln(a_{BX}/a_{+})]$ and $\sim$$\cos^2[s_0 \ln(a_{BX}/a_{+})]$.
For our heteronuclear system, we obtain $s_0 \approx 0.03562$, corresponding to $e^{\pi / s_0} \approx 2\times10^{38}$.
Consequently, the log-period is very large, and the oscillations are inaccessible within any numerically feasible range of $a_{BX}$. Furthermore, the value of $a_{+}$ cannot be reliably determined in our range of $L_3$ calculations, because the Efimov trimer actually does not unbind into the atom-dimer threshold -- as presented in Fig.~\ref{fig:2_merged}(c). Thus, we treat the trigonometric factors as effectively constant. We propose that the interference terms $\sim$$\sin^2[s_0 \ln(a_{BX}/a_{+})]$ and $\sim$$\cos^2[s_0 \ln(a_{BX}/a_{+})]$ are set to 1 and 0, respectively. We emphasize that this approach represents a limiting-case approximation rather than a general prediction. In this way, however, we capture the universal scaling behavior of Efimov-related interferences by introducing a \textit{maximum amplitude} for recombination into a weakly bound Feshbach molecular state, which is given by
 \begin{equation}
      A(L^{\rm w}_3) =  2 C(\delta) \frac{D \left( 1 + \sinh^2\eta \right)}{\sinh^2(\pi s_0 + \eta) } \frac{\hbar a_{BX}^4}{m_X} \ .
      \label{eq:alpha_mod_w}
 \end{equation}
 Similarly, the maximum amplitude for recombination into deep dimers is given by
  \begin{equation}
      A(L^{\rm d}_3) =  2 C(\delta) \frac{D  \coth(\pi s_0) \cosh(\eta) \sinh(\eta) }{\sinh^2(\pi s_0 + \eta)} \frac{\hbar a_{BX}^4}{m_X} \ .
      \label{eq:alpha_mod_d}
 \end{equation}
Naturally, the formula for the total recombination rate is therefore given by
\begin{equation}
    A(L_3) = A(L_3^{\rm w}) + A(L_3^{\rm d})\ .
    \label{eq:A_L_3}
\end{equation}
We fit the model presented in Eq.~\eqref{eq:A_L_3} to our numerical results and estimate the value $\eta \approx 0.027$.

\subsection{Three-body recombination for $a_{BX} < 0$}
For $a_{BX}<0$, shallow, Feshbach dimers are absent. The atoms can only recombine into deep dimers.
Following Ref.~\cite{Helfrich-Hammer:2010}, the rate constant for the recombination into deep dimers is given by
\begin{equation}
          L_3^{\rm d} = {C(\delta)}\frac{D  \coth(\pi s_0)  \sinh(2\eta) }{\sinh^2(\eta) + \sin^2[s_0 \ln(a_{BX}/a_{-})]} \frac{\hbar a_{BX}^4}{m_X} \ ,
      \label{eq:alpha_neg_d}
\end{equation}
where $a_{-}$ is the position of the Efimov resonance. Using limited data for the numerically calculated $L_3$ (obtained up to $|a_{BX}| < 100 \ r_\text{vdW}$) we give a very rough estimate that $a_{-} \approx -200 \ r_\text{vdW}$. Although $|a_{-}| $ is substantially smaller than $a_{+}$, the numerical limit of our calculations does not allow us to confidently include the term $\sin^2[s_0 \ln(a_{BX}/a_{-})]$ in our model for fitting.  Instead, we follow the approach given for $a_{BX} > 0$ and compare our results to the maximum amplitude of the three-body recombination
\begin{equation}
          A(L_3) = {C(\delta)}\frac{D  \coth(\pi s_0)  \sinh(2\eta) }{\sinh^2(\eta) } \frac{\hbar a_{BX}^4}{m_X} \ .
      \label{eq:alpha_neg_d_mod}
\end{equation}
The fit of this model to numerical values allows us to approximate the value $\eta \approx 0.17$.

\bibliography{biblio}

\end{document}